\documentclass[a4paper,11pt]{article}
\usepackage{pos}

\title{The Detection of a Compact Radio Feature in a Seyfert Galaxy After an Accretion Rate Change} \ShortTitle{VLBI observations of KUG\,1141$+$371}

\author*[a,b,c]{Krisztina Éva Gabányi}
\author[d]{Krista Smith}
\author[c,e]{Sándor Frey}
\author[f]{Zsolt Paragi}
\author[g]{Tao An}
\author[c]{Attila Moór}

\renewcommand{\printHeadAuthors}{Gab\'anyi et al.}

\affiliation[a]{Department of Astronomy, Institute of Physics and Astronomy, ELTE Eötvös Loránd University\\
Pázmány Péter sétány 1/A, Budapest, Hungary}

\affiliation[b]{ELKH-ELTE Extragalactic Astrophysics Research Group, ELTE Eötvös Loránd University,\\
Pázmány Péter sétány 1/A, Budapest, Hungary}

\affiliation[c]{Konkoly Observatory, ELKH Research Centre for Astronomy and Earth Sciences (MTA Centre of Excellence), Konkoly-Thege Miklós út 15-17, Budapest, Hungary}

\affiliation[d]{Department of Physics, Southern Methodist University, \\
3215 Daniel Ave, Dallas, TX, USA}

\affiliation[e]{Institute of Physics and Astronomy, ELTE Eötvös Loránd University, \\ Pázmány Péter sétány 1/A, Budapest, Hungary}

\affiliation[f]{Joint Institute for VLBI ERIC, \\
Oude Hoogeveensedijk 4, Dwingeloo, the Netherlands}

\affiliation[g]{Shanghai Astronomical Observatory, Chinese Academy of Sciences, \\
80 Nandan Road, Shanghai, People's Republic of China} 

\emailAdd{k.gabanyi@astro.elte.hu}
\emailAdd{kristas@smu.edu}
\emailAdd{frey.sandor@csfk.org}
\emailAdd{zparagi@jive.eu}
\emailAdd{antao@shao.ac.cn}
\emailAdd{moor.attila@csfk.org}

\abstract{X-ray binaries are known to show state transitions related to accretion rate changes which are often accompanied with dramatic changes in the jet emission. However, it is not clear whether this characteristics of stellar-mass black hole systems can be scaled up to the accretion disk of active galactic nuclei.
The Seyfert 1 galaxy, KUG\,1141$+$371 has been showing a steadily increasing X-ray flux since 2007, and exhibited variability behaviour similar to the state transitions observed in X-ray binaries. It was hypothesised to undergo a rapid boost of mass accretion. If the X-ray binary analogy holds then the appearance of jet emission can also be expected in KUG\,1141$+$371. While the source was not detected in the Faint Images of the Radio Sky at Twenty-centimeters in 1994, it appears in the VLA Sky Survey in 2019 and at $22$\,GHz in a VLA observation in 2018 at mJy flux density level. Our VLBI observations revealed a compact, flat-spectrum radio feature. Its high brightness temperature indicates the radio emission originates from an AGN.}

\FullConference{%
  15th European VLBI Network Mini-Symposium and Users' Meeting (EVN2022)\\
  11-15 July 2022\\
  University College Cork, Ireland
}

\begin{document}
\maketitle


\section{Introduction}

Variability of the optical and X-ray emissions from the nuclei of Seyfert galaxies on timescales shorter than the viscous timescale in a standard thin-disk model is usually explained by either variations in obscuration, changes in bulk accretion rate, or turbulent processes within the accretion disk. In stellar-mass black hole systems, the complex variability of the X-ray luminosity and the hardness ratio is explained by changes in the accretion disk structure. In microquasars, synchrotron self-absorbed core-jets are ubiquitous in the hard state and they are responsible for the X-ray--radio correlation in such systems \cite{fender_science}. However, it is still unclear whether and how the disk--corona system of X-ray binaries can be scaled up to describe the optical-UV emissions of the accretion disks of active galactic nuclei (AGN). 

Jiang et al. \cite{jiang} described the multi-wavelength variability observed in the Seyfert 1 galaxy, KUG 1141$+$371 ($z=0.038$). Notably, its hard X-ray flux has increased by over a magnitude since 2009, while its soft X-ray flux has also shown a steady increase since 2007. They used the optical and UV data to describe the thermal emission of the accretion disk, and calculated the X-ray-to-UV ratio to use it analogously to the X-ray hardness of stellar-mass black hole systems. From the modeling of the spectral energy distribution (SED), they argue that the variability behaviour of KUG 1141$+$371 is similar to that of the state transitions seen in X-ray binaries and can be related to changes in the accretion state. The steady increase of the X-ray flux and broad-band SED suggest a rapid boost of mass accretion rate in KUG 1141$+$371 that could also produce simultaneous radio variability if the stellar-mass black hole analogy is correct. While the changes in the optical--X-ray spectral index and in the Eddington accretion rate is similar to changing-look AGN, KUG 1141$+$371 has not changed ``look'' so far but retains a Seyfert 1 galaxy spectrum since its original classification in 2005 \cite{classification}. KUG 1141$+$371 was not detected in the Faint Images of the Radio Sky at Twenty-centimeters (FIRST, \cite{first}) in 1994 (thus with a flux density $<1$\,mJy), but it appears in the Very Large Array Sky Survey (VLASS, \cite{vlass}) in 2019 in the frequency range of $2-4$\,GHz, with a flux density of $(1.14\pm0.25)$\,mJy. Moreover, it was detected by the VLA in C configuration at $22$\,GHz (proposal id: 18B-245, PI: K. Smith) on 2018 Nov 27 with a flux density of $(0.53\pm0.02)$\,mJy. Here we present the results of our very long baseline interferometry (VLBI) observations of KUG\,1141$+$371. 


\section{Observations and Data Reduction}

KUG\,1141$+$371 was observed with the joint array of the European VLBI Network and the enhanced Multi-Element Remotely Linked Interferometer Network (EVN+e-MERLIN) at $1.7$ and at $4.9$\,GHz on 2021 June 4 and May 11, respectively (project code: EG116). The following antennas provided data at both frequencies: Jodrell Bank (United Kindom), Effelsberg (Germany), Onsala (Sweden), Svetloe, Badary (Russia), Irbene(Latvia), Westerbork (The Netherlands), and four antennas of the e-MERLIN array, Cambridge, Darnhall, Knockin, and Pickmere (United Kingdom). Medicina, Noto, Sardinia (Italy), Toru\'n (Poland), Tianma, Urumqi (China), and Zelenchukskaya (Russia) also observed at $1.7$\,GHz; and Shanghai (China), Defford and the Lovell Telescope of the e-MERLIN array at $4.9$\,GHz.

At $1.7$\,GHz, four intermediate frequencies (IFs) were used, each with $64$ spectral channels and $2$\,s of integration time, except for the e-MERLIN antennas where observation was done in only IF $2$ and $3$. At $4.9$\,GHz, $8$ IFs, each with $64$ spectral channels and $1$\,s of integration time were used. However, Badary, Svetloe, Shanghai, and Westerbork observed only in the lower $4$ IFs, and the e-MERLIN antennas only in IF $6$ and $7$  due to bandwidth limitations. At both frequencies, $4$ polarizations were recorded, but at $1.7$\,GHz Westerbork provided data in left circular polarization only. The observations lasted for $3$ and $4$ hours at $1.7$ and $4.9$\,GHz, respectively. The on-source times were $1.7$\,h at $1.7$\,GHz, and $3.6$\,h at $4.9$\,GHz. The observations were conducted in phase-reference mode with ICRF\,J114933.9$+$355908 as the phase-reference calibrator separated from our target by $1.36^\circ$.

A priori flagging, amplitude and phase calibrations were done in the Astronomical Image Processing System (AIPS, \cite{aips}) following standard procedures. Then the fringe-fitted phase reference source was imaged with the Difmap software \cite{difmap} using several iterations of cleaning \cite{clean} and phase self-calibration. The result of the imaging was used to determine amplitude correction factors which were then applied in AIPS to the visibility amplitudes. Then fringe-fitting was performed again for the phase-reference source taking into account its structure. The obtained solutions were applied to the target source, then its data were exported into Difmap for imaging. No self-calibration was attempted for the target source. 

\section{Results}

\begin{figure}
\centering
\includegraphics[bb=0 90 720 450, clip=,width=\textwidth]{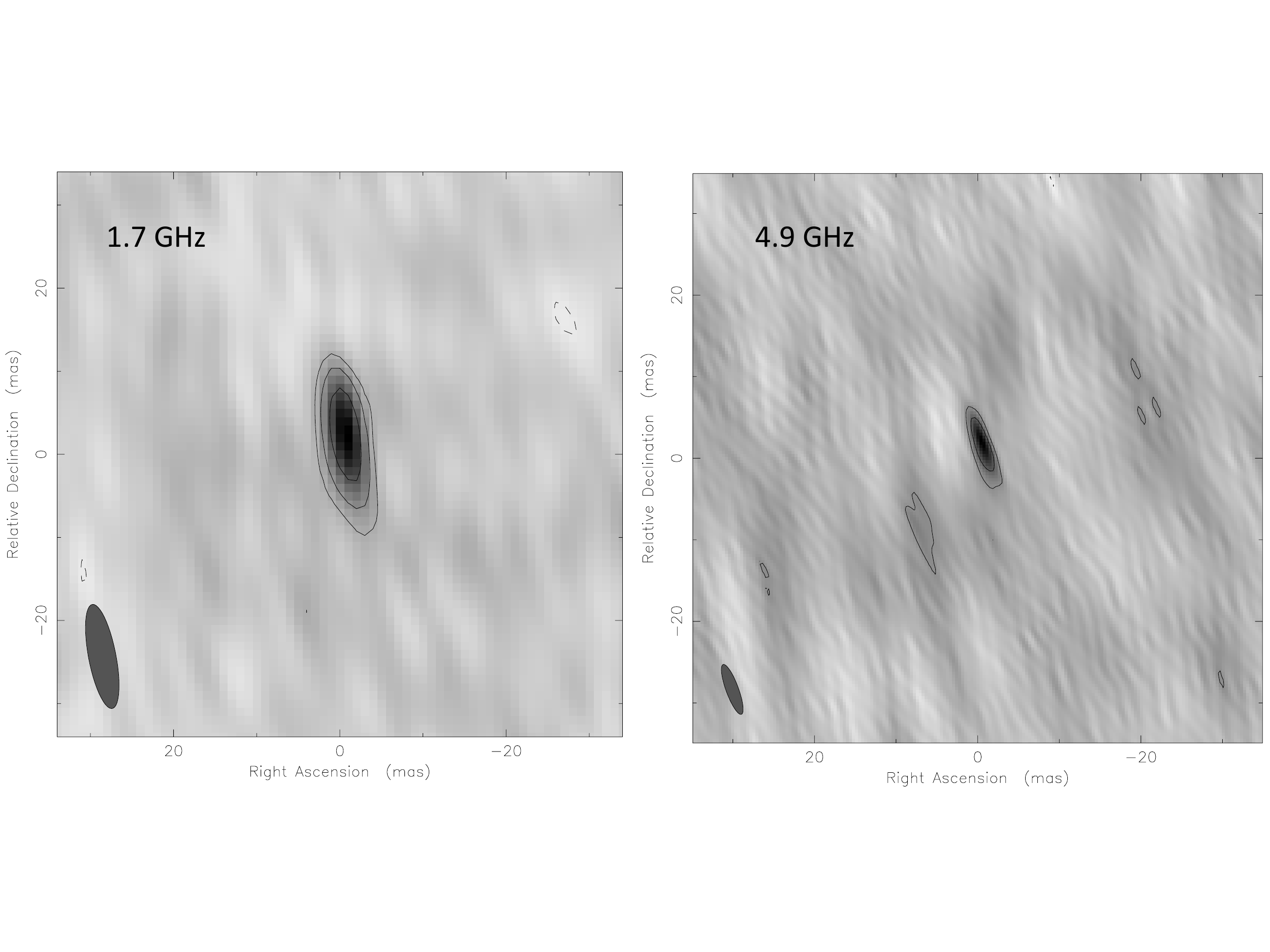}
\caption{EVN+e-MERLIN images of KUG\,1141$+$371. {\it Left panel:} $1.7$-GHz VLBI image. Peak intensity is $0.25\textrm{\,mJy\,beam}^{-1}$. Lowest positive contour is at $0.036\textrm{\,mJy\,beam}^{-1}$, at $3\sigma$ image noise level, further contour levels increase by the factor of two. Restoring beam is $12.8 \textrm{\,mas} \times 3.3\textrm{\,mas}$ at a major axis position angle of $10.9^\circ$ and it is shown at the lower left of the image. {\it Right panel:} $4.9$-GHz VLBI image. Peak intensity is $0.21\textrm{\,mJy\,beam}^{-1}$. Lowest positive contour is at $0.053\textrm{\,mJy\,beam}^{-1}$, at $3\sigma$ image noise level, further contour levels increase by the factor of two. Restoring beam is $6.5 \textrm{\,mas} \times 1.35\textrm{\,mas}$ at a major axis position angle of $20.2^\circ$ and it is shown at the lower left of the image.  \label{fig:EVNmaps}}
\vspace{-0.3cm}
\end{figure}

\begin{figure}
\centering
\includegraphics[bb=20 30 640 470, width=0.6\textwidth,clip=]{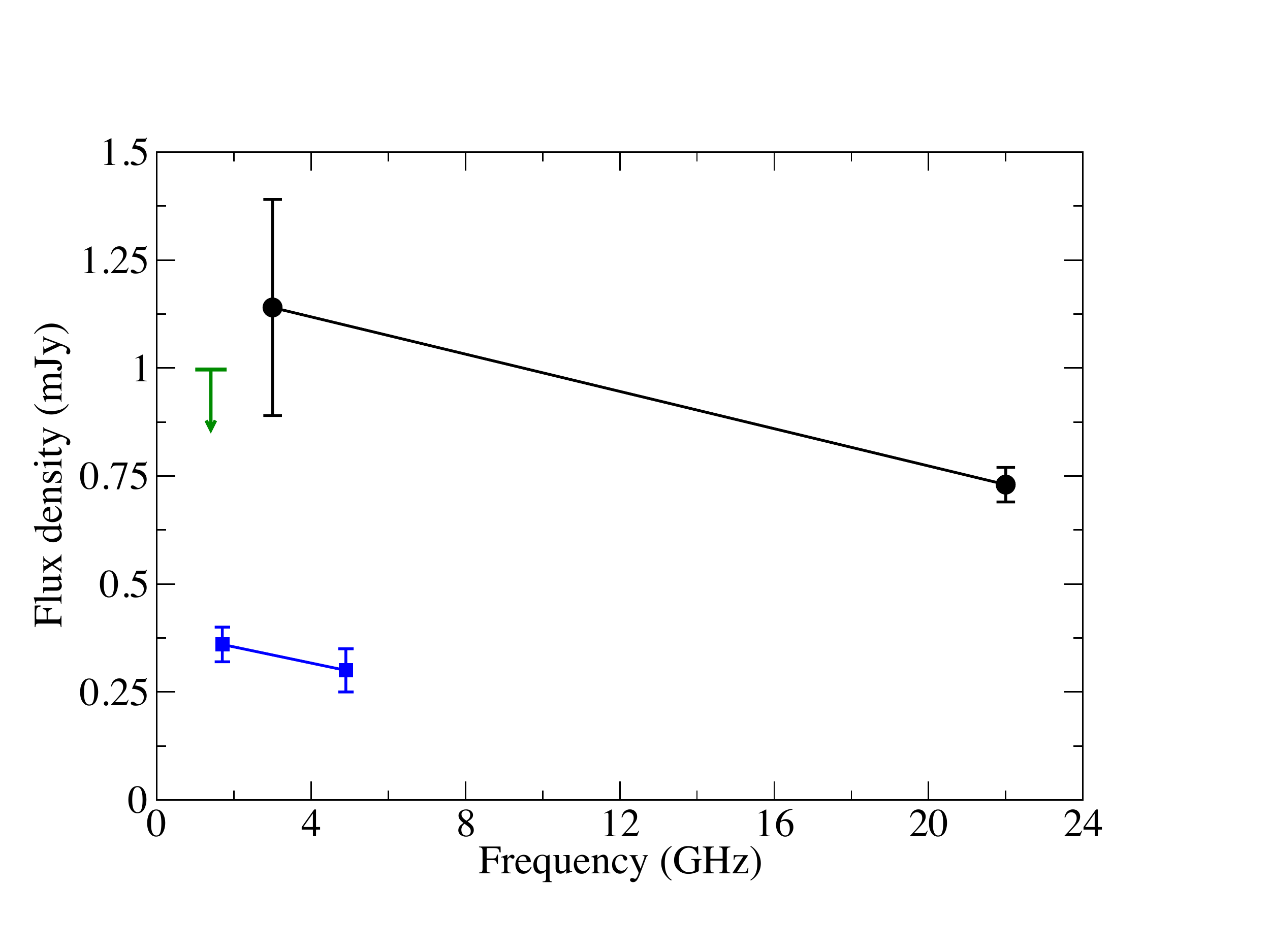}
\caption{Radio spectrum of KUG\,1141$+$371. Blue, and black symbols represent VLBI, and VLA flux densities. The latter are from the VLASS \citep{vlass} and from the VLA project, 18B-245. Green arrow shows the FIRST upper limit \cite{first}. \label{fig:radiospectrum}}
\vspace{-0.3cm}
\end{figure}

At both frequencies, KUG\,1141$+$371 appeared as a single compact component (Fig.\,\ref{fig:EVNmaps}). The coordinates of the brightest pixel in the $4.9$\,GHz image are right ascension $\alpha=11^\textrm{h}44^\textrm{m} 29.87078^\textrm{s}$ and declination $\delta=36^\circ 53' 8.616711''$ with an accuracy of $\sim1$\,mas \cite{chatterjee}. These coordinates agree within the error with the {\it Gaia} DR3 optical coordinates of KUG\,1141$+$371 \cite{gaiadr3}.

To describe the brightness distribution of KUG\,1141$+$371, we fitted the visibilities with 2-dimensional, circular Gaussian function. For deriving the flux densities, we assumed a coherence loss of $\sim 25$\,\% \cite{coherence1}. 
Thus, the flux densities are $S_{1.7}=(0.36\pm0.04)$\,mJy, and $S_{4.9}=(0.3\pm0.1)$\,mJy. Assuming a flat $\Lambda$CDM cosmology with $H_0=70 \textrm{\,km\,s}^{-1}\textrm{\,Mpc}^{-1}$, $\Omega_\Lambda=0.73$, and $\Omega_\textrm{m}=0.27$, the corresponding radio powers are $P_{1.7}=(1.2 \pm 0.1)\cdot10^{21}\textrm{W\,Hz}^{-1}$ and $P_{4.9}=(1.0 \pm 0.2)\cdot10^{21}\textrm{W\,Hz}^{-1}$.
The full-width half-maximum (FWHM) size at $1.7$\,GHz is $(2.2\pm0.1)$\,mas. At $4.9$\,GHz, the FWHM size of the fitted component is below the smallest resolvable size, $1.81$\,mas \cite{kovalev_size}. Therefore, the brightness temperature can only be calculated at $1.7$\,GHz, $T_\textrm{B,1.7}=3.3\cdot10^7$\,K, while only a lower limit can be given at $4.9$\,GHz, $T_\textrm{B, 4.9}\gtrsim 1.1 \cdot10^{8}$\,K. These high values indicate that the radio emission in  KUG\,1141$+$371 originates from an AGN \cite{condon_gal}. Using the flux densities of the fitted Gaussian components, the spectral index, $\alpha$ (defined as $S\sim\nu^\alpha$, where $\nu$ is the observing frequency) can be calculated. The obtained value, $\alpha_\textrm{VLBI}=-0.2\pm0.3$ indicates a flat radio spectrum. The low-resolution observations with the VLA indicate a similar flat spectrum with a spectral index of $\alpha_\textrm{VLA}=-0.2\pm0.1$ (Fig. \ref{fig:radiospectrum}). On the other hand, the VLBI flux densities are well below the VLA values indicating, either significant arcsec-scale emission and/or variability on a timescale of $2-3$ years. We calculated the ratio of the X-ray and radio luminosity using the latest X-ray measurement from $2019$ \cite{jiang} and our $5$\,GHz EVN data. The ratio, $\sim1.3\times10^{-7}$, is two orders of magnitude lower than the one expected from the G\"udel--Benz relation assuming an accretion corona-related radio emission \citep{laor2008}. However, the radio and X-ray measurements are not simultaneous. Further X-ray and radio monitoring can provide information on the origin of the radio emission.

\section*{Acknowledgements}

The European VLBI Network is a joint facility of independent European, African, Asian, and North American radio astronomy institutes. e-MERLIN is a National Facility operated by the University of Manchester at Jodrell Bank Observatory on behalf of STFC. Scientific results from data presented in this publication are derived from the following EVN project code: EG116. The research leading to these results has received funding from the European Union's Horizon 2020 Research and Innovation Programme under grant agreement No.\,101004719 (OPTICON RadioNet Pilot). This research was supported by the Hungarian National Research, Development and Innovation Office (OTKA K134213).

\end{document}